\documentclass[pre,superscriptaddress,tightenlines,twocolumn,showpacs]{revtex4-1}
\pdfoutput=1
\usepackage{ulem}

\usepackage{color}
\definecolor{nblue}  {RGB}{28,130,185}

\definecolor{cgreen}  {RGB}{76,153,0}

\usepackage{latexsym,amsmath,amsfonts,tabularx,amssymb,bm,empheq}
\usepackage{graphicx,epstopdf}
\usepackage{xspace}

\usepackage{hyperref}

\newcommand{\bea}{\begin{eqnarray}}
\newcommand{\eea}{\end{eqnarray}}

\newcommand\subrel[2]{\mathrel{\mathop{#2}\limits_{#1}}}

\newcommand{\dfonc}[2]{\frac{\delta#1}{\delta#2}}
\newcommand{\dd}{\mathrm{d}}
\newcommand{\ddd}{\mathcal{D}}
\newcommand{\ee}{\mathrm{e}}
\newcommand{\phit}{\tilde{\phi}}

\newcommand{\psit}{\tilde{\psi}}

\newcommand{\bra}{\left\langle}
\newcommand{\ket}{\right\rangle}

\def\simge{\mathrel{%
   \rlap{\raise 0.511ex \hbox{$>$}}{\lower 0.511ex \hbox{$\sim$}}}}
\def\simle{\mathrel{
   \rlap{\raise 0.511ex \hbox{$<$}}{\lower 0.511ex \hbox{$\sim$}}}}

\def\simle{\mathrel{
   \rlap{\raise 0.511ex \hbox{$<$}}{\lower 0.511ex \hbox{$\sim$}}}}

\def\simge{\mathrel{%
    \rlap{\raise 0.511ex \hbox{$>$}}{\lower 0.511ex \hbox{$\sim$}}}}

\usepackage{subfigure}
\usepackage{color}

\usepackage{tabularx}

\begin{document}

\title{Frequency regulators for the nonperturbative renormalization group: \\
A general study and the model A as a benchmark}

\author{Charlie Duclut and Bertrand Delamotte}
%\affiliation{
%Laboratoire de Physique 
%Th\'eorique de la Mati\`ere Condens\'ee, CNRS UMR7600, Universit\'e Pierre et Marie Curie, 
%75252 Paris Cedex 05, France}
\affiliation{
Laboratoire de Physique Th\'eorique de la Mati\`ere Condens\'ee, UPMC,
CNRS UMR 7600, Sorbonne Universit\'es, 4, place Jussieu, 75252 Paris Cedex 05, France
}

\date{\today}

\begin{abstract}
We derive the necessary conditions for implementing a regulator that depends on both momentum and frequency in the nonperturbative renormalization group flow equations of out-of-equilibrium statistical systems. We consider model A as a benchmark and compute its dynamical critical exponent $z$. This allows us to show that frequency regulators compatible with causality and the fluctuation-dissipation theorem can be devised.
We show that when the Principle of Minimal Sensitivity (PMS) is employed to optimize the critical exponents $\eta$, $\nu$, and $z$, the use of frequency regulators becomes necessary to make the PMS a self-consistent criterion. 
\end{abstract}

\maketitle

\section{Introduction}

Non-equilibrium statistical physics and in particular the study of out-of-equilibrium phase transitions has become in the past few decades one of the challenges of statistical physics. A large variety of tools have been developed to tackle these problems, and one of the most powerful is the dynamical Renormalization
Group~(RG)~\cite{Ma_book,Tauber_book}, which extends the equilibrium field-theoretic approach to non-equilibrium. Although many basics features of the analog of the partition function (its positivity, its convexity with respect to sources, etc.) have still not been proven in general, many results have been achieved using dynamical RG techniques, for instance the description and characterization of several universality classes~\cite{Cardy98,Hinrichsen2000,Janssen05}. 

The nonperturbative and functional version of this approach has also proven very well-suited for non-equilibrium problems~\cite{Canet11,*Canet12}, probably because many of them present intrinsically non-perturbative characteristics that prevent the usual RG from being effective in these cases. For instance, some models yield exact results within the RG formalism at all orders in the perturbative expansion, although these results are incomplete, that is, they fail to account for the corresponding experimental, numerical or exact results~\cite{Janssen04,Wiese98,Canet05_2}. This seems to indicate some non-analytical features of the models that only a non-perturbative and/or functional approach could handle correctly~\cite{Canet04,Canet05_2,Canet06,Canet10,Canet11_2,Kloss14,Kloss14_2} (see also \cite{Tarjus08,Tissier06,*Tissier08,*Tissier12a,*Tissier12b} for the same kind of problems in equilibrium disordered systems).

The key to the success of the nonperturbative renormalization group (NPRG) approach in equilibrium physics is its ability to take care of growing fluctuations near criticality  by integrating them out in a controlled way~\cite{Berges02}; for an introduction, see~\cite{Gies12,Delamotte12,Delamotte05}. This is achieved by coarse-graining the spatial fluctuations using a regulator function $R_k(\vert x-y\vert)$ in the action of the model which has a typical range $\vert q\vert\lesssim k$ in momentum space. This key feature of the NPRG, reminiscent of the block-spin idea, is probably not sufficient in many non-equilibrium problems, where temporal fluctuations also play a major role. The introduction of a regulator that would also take care of these temporal fluctuations therefore seems important. A first example in which a frequency regulator could be needed is the Parity-Conserving-Generalized-Voter model,
which is a one-species reaction-diffusion system where the parity of the number of particles is conserved by the dynamics~\cite{Kockelkoren03,Dornic01}. 
Some approximate results obtained with the NPRG~\cite{Canet05_2} for this model disagree qualitatively with exact ones~\cite{Benitez13}, indicating that the fluctuations are not properly taken into account, at least within the Local Potential Approximation (LPA), which has proven to be very efficient in equilibrium problems. 

To be more specific, the most used approximation in the NPRG context is the Derivative Expansion (DE)~\cite{Morris94,Berges02}. In this approximation, the contributions of all the correlation functions to the RG flow  are retained but their momentum/frequency dependence is replaced by a Taylor expansion. The DE is valid and accurate if the  radius of convergence of this Taylor expansion is larger than the range of the momentum and frequency integrals in the RG flow equations. In this case, the correlation functions and the propagator used in the RG flow are well approximated by their momentum/frequency expansion, at least in the region that contributes to the integrals in the flow, that is, in the region that is not suppressed by the regulator $R_k(q)$.  

At equilibrium, the role of the regulator $R_k(q)$ introduced within the NPRG framework is therefore to effectively cut off the momentum integration from  $\vert q\vert\in [0,\infty[$ to $0\le \vert q\vert\lesssim k$.  The radius of convergence of the momentum-expansion of the correlation functions is probably of order $k$ which coincides with the typical range of the regulator $R_k(q)$. This allows for the replacement of the correlation functions and the propagator by their Taylor expansion in the integrals of the flow equations, and it probably explains the success of the DE~\cite{Morris99,Benitez12}. For nonequilibrium systems,  this issue is subtler because the RG flow equations involve also a frequency integral. This integral is convergent~\footnote{This is true at order two of the DE, but depending on the approximations performed, the integral over the frequencies can diverge, in which case regularization is of course necessary.} without any regularization  which means that the integrand decreases sufficiently rapidly for the region of large frequencies to contribute a finite amount.
However, the fact that the frequency integral is convergent does not guarantee that it is accurately computed when the  correlation functions are replaced in the integrand by their  frequency-expansion.
Therefore, this integral must also be cut off by a regulator to avoid summing contributions at large frequencies corresponding to a region where the Taylor expansion of the correlation functions is not valid.

Our goal here is to design  frequency regulators that  generalize the role played by the regulators in the usual equilibrium NPRG settings to non-equilibrium cases. So far, and to the best of our knowledge, such a regulator has not been engineered, and we discuss here the theoretical properties that it should show in order to both fulfill its regularization role and enforce important physical constraints such as causality and the fluctuation-dissipation theorem.

The model A, also called the kinetic Ising model, and its multidimensional-spin counterpart (the kinetic $O(N)$ model) \cite{Hohenberg77} are used as benchmark models to test our regulators.

\section{Model A and its field-theoretic formulation}
		
The model A or kinetic Ising model is one of the simplest models one can think of to describe out-of-equilibrium critical phenomena. It is a coarse-grained description of Glauber dynamics for Ising spins \cite{Tauber_book,Hohenberg77}~: On a $d$-dimensional lattice, Ising spins are allowed to flip with transition rates that depend on the orientation of their neighbors and satisfy the detailed balance condition, guaranteeing the system will relax towards equilibrium at large time. The model A uses a Langevin description of the spins dynamics, and it is therefore stated in terms of a coarse-grained local spin variable $\phi (x,t)$ following the stochastic equation (in the It\= o sense):
\begin{align}
	\partial _t \phi (x,t) = - \frac{\mathrm{\delta}  H}{\mathrm{\delta}  \phi} + \zeta (x,t) \, , \label{eq_Langevin}
\end{align}
where $H=H[\phi]$ is the usual Ginzburg-Landau Hamiltonian:
\begin{align}
	H[\phi] = \int _{\bm{x}} \left( \frac{1}{2} \left( \nabla \phi \right) ^2 + V(\phi)\right)
\end{align}
with $\bm x \equiv (x,t)$,
$\int_{\bm x} \equiv \int \dd^d x \,\dd t $,
$V(\phi)=r/2 \, \phi ^2 + u/4! \, \phi ^4$, and $\zeta (\bm x)$ is a Gaussian white noise taking into account the fluctuations of the order parameter coming from its coarse-grained nature. The noise probability distribution $P(\zeta )$ is consequently:
\begin{align}
	P(\zeta ) \propto \ee^{-\frac{1}{4}\int _{\bm x} \zeta (\bm x) ^2}
\end{align}
yielding in particular
\begin{align}
	\bra \zeta (\bm x) \zeta (\bm{x'}) \ket = 2\, \delta (t-t') \delta  ^d (x-x') \, ,
\end{align}
where we have rescaled the time and the Hamiltonian such that the variance of the noise is 2. From this Langevin equation a field-theoretic approach can be derived using the Martin-Siggia-Rose-de Dominicis-Janssen (MSRDJ) approach \cite{Martin73,Janssen76,DeDominicis76}. In this formalism, the mean value (over the realizations of the noise) of a given observable $\mathcal{O}[\phi]$ is given by:
\begin{align}
	\bra \mathcal{O}[\phi ] \ket _{\zeta} &=\int \ddd \phi  \ddd \tilde{\phi} \,	\ee ^{-\mathcal{S}[\phi,\tilde{\phi}]} \mathcal{O}[\phi] \label{eq_sansJacob}
\end{align}
with 
\begin{align}
    \mathcal{S}[\phi,\tilde{\phi}] = \int _{\bm x} \tilde{\phi} 
        \left( \partial_t\phi - \tilde{\phi} + \frac{\mathrm{\delta}H}{\mathrm{\delta}\phi}
        \right) \, .
        \label{eq:action}
\end{align}
Notice that within this formalism the functional integral over $\phit$ (which is called the ``response'' field for reasons that will become clear in the following) is performed along the imaginary axis, whereas $\phi$ is a real field.

\section{Non-Perturbative RG formalism}

As in equilibrium statistical physics, the starting point of the field theory is the analog of the partition function associated with the previous action $\mathcal S$ defined in Eq.~(\ref{eq:action}), and which reads:
\begin{align}
    \mathcal{Z}[j,\tilde{j}] = \int \mathcal{D} \phi  \mathcal{D}\tilde{\phi} \, 
    \mathrm{e}^{-\mathcal{S}+\int_{\bm x}J(\bm x)^T \cdot \Phi (\bm x)}
\end{align}
where we use a matrix notation and define the following vectors
\begin{align}
    \Phi (\bm x)= 
        \left( 
             \begin{array}{c} \phi(\bm x) \\ \tilde{\phi}(\bm x)  \end{array} 
        \right) 
    \quad \text{and} \quad
    J(\bm x)= 
        \left( 
             \begin{array}{c} j(\bm x) \\ \tilde{j}(\bm x)  \end{array} 
        \right) \, .
\end{align}
As in equilibrium, the generating functional of the connected correlation and response functions is  $\mathcal{W}[J] = \log \mathcal{Z}[J]$. We also introduce its Legendre transform, the generating functional of the one-particle irreducible correlation functions $\Gamma[\Psi]$, where $\Psi= \bra \Phi \ket$.

In order to determine the effective action $\Gamma$, we apply the NPRG formalism and write a functional differential equation which interpolates between the microscopic action~$\mathcal{S}$ and the effective action~$\Gamma$. The interpolation is performed through a momentum scale $k$ and by integrating over all the fluctuations with momenta $\vert q\vert>k$, while those with momenta $\vert q\vert<k$ are frozen. At scale $k=\Lambda$, where $\Lambda$ is the ultra-violet cutoff imposed by the (inverse) microscopic scale of the model (e.g. the lattice spacing), all fluctuations are frozen and the mean-field approximation becomes exact; at scale $k \to 0$, all the fluctuations are integrated over and the original functional $\mathcal{Z}$ is recovered. The interpolation between these scales is made possible by using a regulator $\mathcal{R}_k(\bm x)$, whose role is to freeze out all the fluctuations with momenta $\vert q\vert<k$. This regulator is introduced by adding an extra term to the action and thus defining a new partition function $\mathcal Z_k$:
\begin{align}
    \mathcal{Z}_k[j,\tilde{j}] = \int \mathcal{D} \phi  \mathcal{D}\tilde{\phi} \, 
    \mathrm{e}^{-\mathcal{S}-\Delta \mathcal S_k+\int_{\bm x}J(\bm x)^T \cdot \Phi (\bm x)}
\end{align}
with
\begin{align}
    \Delta\mathcal{S}_k= \frac{1}{2}\int_{\bm x,\bm{x'}} \Phi(\bm x)^T \cdot \mathcal{R}_k(\bm x-\bm{x'}) \cdot \Phi (\bm{x'})
    \label{eq:def_regulator}
\end{align}
where $\mathcal{R}_k$ is a $2\times 2$ regulator matrix, depending both on space and time, and whose task is to cancel slow-mode fluctuations. We shall see in the following sections that the regulator form (and especially its frequency part) is constrained by causality and by symmetry considerations.
We also define the effective average action $\Gamma_k$ as a modified Legendre transform  of $\mathcal{W}_k[J] = \log \mathcal{Z}_k[J]$~\cite{Wetterich93}:
\begin{align}
\begin{split}
    &\Gamma_k[\Psi]+\mathcal{W}_k[J] = \\
    &\int_{\bm x} J^T\cdot \Psi - \frac{1}{2} \int_{\bm x,\bm{x'}} \Psi(\bm x)^T \cdot \mathcal{R}_k(\bm x-\bm{x'}) \cdot \Psi (\bm{x'})
\end{split}
\end{align}
in such a way that $\Gamma_k$ coincides with the action at the microscopic  scale -- $\Gamma_{k=\Lambda}=\mathcal S$ -- and with $\Gamma$ at $k=0$ --~$\Gamma_{k=0}=\Gamma$--  , that is, when all fluctuations have been integrated over. The evolution of the interpolating functional $\Gamma_k$ between these two scales is given by the Wetterich equation~\cite{Wetterich93,Morris94_2}:
\begin{align}
    \partial_k \Gamma_k [\Psi] &= \frac{1}{2} \text{Tr} \int_{\bm x,\bm{x'}} \partial_k \mathcal{R}_k(\bm x-\bm{x'}) \cdot G_k [\bm x,\bm{x'};\Psi]
    \label{eq:Wetterich}
\end{align}
where $G_k [\bm x,\bm{x'};\Psi] \equiv [ \Gamma_k^{(2)}+\mathcal{R}_k]^{-1}$ is the full, field-dependent, propagator and $\Gamma_k^{(2)}$ is the $2 \times 2$ matrix whose elements are the $\Gamma_{k,ij}^{(2)}$ defined such that:
\begin{align}
    \Gamma_{k,i_1,\cdots,i_n}^{(n)}[{\bm x_i};\Psi] &= \frac{\delta^n \Gamma_k[\Psi]}{\delta \Psi_{i_1}(\bm x_1)\cdots \delta \Psi_{i_n}(\bm x_n)} \, .
\end{align}
The Wetterich equation~(\ref{eq:Wetterich}) represents an exact flow equation for the effective average action $\Gamma_k$, which we solve approximately by restricting its functional form. We use in the following the DE, stating that instead of following the full $\Gamma_k$ along the flow, only the first terms of its series expansion in space and time derivatives of $\Psi$ are considered. These terms have to be consistent with the symmetries of the action~$\mathcal S$, and we therefore discuss them before giving an explicit \textit{ansatz} for $\Gamma_k$.

Since the model A should relax towards thermodynamic equilibrium at large time, one expects the partition function~$\mathcal Z$ to be symmetric under time reversal, and this leads to the invariance of the action~$\mathcal S$ under the following transformation~\cite{Andreanov06,Aron10}:
\begin{align}
\begin{cases}
    \phi' (x,t) & \hspace{-3mm}= \phi (x,-t) \\
    \tilde{\phi}'(x,t) &\hspace{-3mm}=  \tilde{\phi}(x,-t)-\dot\phi(x,-t)
\end{cases}
\label{eq_invariance}
\end{align}
where $\dot f(t) \equiv \partial_t f(t)$.  Provided that the noise term is Gaussian, the invariance of the action under this transformation is the signature of the equilibrium dynamics and of the Fluctuation-Dissipation Theorem~(FDT)~\cite{Aron10}. 

The previous considerations about FDT allow us to write the following \textit{ansatz} for $\Gamma_k$, at first order in time derivative, and second order in space derivative~\cite{Canet11}:
\begin{align}
    \Gamma_k[\psi,\tilde{\psi}]
    &= \int_{\bm x} \tilde{\psi}\left[ X_k(\psi)\left( \partial_t \psi -\tilde{\psi}\right) + \dfonc{\gamma_{{\rm eq},k}}{\psi({\bm x})} \right] \nonumber \\
    \begin{split}
        &=\int_{\bm x} \tilde{\psi}\left[ X_k(\psi)\left( \partial_t \psi -\tilde{\psi} \right) + U_k'(\psi)  \right. \\
        & \left. - Z_k(\psi) \nabla^2 \psi - \frac{1}{2} Z_k'(\psi)(\nabla \psi)^2   \right] \label{eq:derivativeExpansion}
    \end{split}
\end{align}
where, at equilibrium:
\begin{align}
   \Gamma_{{\rm eq},k}[\psi] &=\int \dd^d x\,\gamma_{{\rm eq},k}(\psi(x,t))\\
   &=\int \dd^d x\,\left[ \frac{1}{2} Z_k(\psi)(\nabla \psi)^2 + U_k(\psi) \right].
\end{align}
Let us briefly justify the form of $\Gamma_k$. It
is natural to choose it invariant under transformation (\ref{eq_invariance}) so that FDT holds at all $k$. 
This implies that the terms proportional to $\tilde{\psi}^2$ and $\tilde{\psi}\partial_t\psi$  renormalize in the same way and therefore depends on a single function $X_k$, which is a tremendous simplification of the RG flow. The second part of the \textit{ansatz}, $\tilde\psi \, \partial \gamma_{{\rm eq},k}/\partial\psi(x,t)$, is linear in $\tilde{\psi}$ and is therefore invariant on its own since the transformation (\ref{eq_invariance}) generates a term proportional to $\partial_t \psi \, \partial \gamma_{{\rm eq},k}/\partial \psi(x,t)=\partial_t \gamma_{{\rm eq},k}(\psi)$ that vanishes after time integration in the stationary regime. Notice that because of the FDT, higher-order terms in $\tilde\psi$ are not allowed at this order of the DE, and thus $U_k$,  $Z_k$, and $X_k$ are functions of $\psi$ only (see \cite{Canet07} for further explanations).

Choosing \textit{ansatz}~(\ref{eq:derivativeExpansion}) implies to use only regulators compatible with (\ref{eq_invariance}) and we show in the following that it is indeed possible to devise such regulators even when they depend on frequencies. Of course, it is possible to consider regulators that are incompatible with (\ref{eq_invariance}) at the price of giving up FDT for $k>0$. This implies that in $\Gamma_k$ the two terms $\tilde{\psi}^2$ and $\tilde{\psi}\partial_t\psi$ do no longer renormalize in the same way. In this case, the \textit{ansatz}~(\ref{eq:derivativeExpansion}) becomes
\begin{align}
    \Gamma_k[\psi,\tilde{\psi}]
    &= \int_{\bm x} \tilde{\psi}\left[ X_k(\psi) \partial_t \psi + W_k (\psi) \tilde{\psi} + \dfonc{\gamma_{{\rm eq},k}}{\psi({\bm x})} \right]\, .
\end{align}
Notice that when the field dependence of $X_k(\psi)$ and $W_k(\psi)$ is neglected ($X_k(\psi)\to\bar X_k$ and $W_k(\psi)\to\bar W_k$)
and that the regulator is frequency-independent the flows of $\bar X_k$ and $\bar W_k$ are identical~\cite{Chiocchetta16}. This incidental property is however lost when the  field dependence of these functions is kept.

Using \textit{ansatz}~(\ref{eq:derivativeExpansion}) drastically simplifies the resolution of the Wetterich equation since a functional differential equation is converted into a set of partial differential equations over the functions $U_k$,  $Z_k$ and $X_k$. The role of the regulator is essential for the validity of this approximation, and we therefore discuss its properties in more detail in the following section.

\section{Frequency regulator}
\label{sect:regu}

Now that we have introduced the NPRG formalism, and before explaining how the flow of the different renormalization functions is computed, we have to focus on the regulator term, whose role is crucial for ensuring the validity of the DE and the stability of the form of the \textit{ansatz} (\ref{eq:derivativeExpansion}) under the RG flow.

Let us first remind that the MSRDJ formalism together with Ito's prescription does not allow for a term in the action not proportional to the response field $\tilde\phi$. This implies that there is no cut-off term in the $\phi ^2$ direction, and  the regulator matrix defined in Eq.~(\ref{eq:def_regulator}) can be written in full generality as:
\begin{align}
    \mathcal{R}_k(\bm x)=\left( 
         \begin{array}{cc} 0 & R_{1,k}( x,t) \\ R_{1,k}( x,-t) &  2 R_{2,k}( x,t)  \end{array} 
    \right) 
\end{align}
where the minus sign in $R_{1,k}(x,-t)$ is a consequence of $\Delta \mathcal{S}_k$ being written in a matrix form and the factor 2 in front of $R_{2,k}$ has been included for convenience. Notice that these two regulator terms have a meaning for the underlying Langevin equation. Indeed, adding a regulator $R_{1,k}$ means changing the external force in the Langevin equation:
\begin{align}
    F \equiv -\dfonc{H}{\phi} \, \rightarrow \, F + \Delta F_k%- \int_{\bm u} R_{1,k} (\bm u-\bm x) \phi(\bm u)
\end{align}
where $\Delta F_k({\bm x}) = -\int_{\bm u} R_{1,k}(\bm u-\bm x) \phi(\bm u)$. The regulator $R_{1,k}$ is thus similar to the usual mass-like regulator used at equilibrium.
We restrict ourselves in the following to  additional forces $\Delta F_k$ which are causal. This implies $R_{1,k}(x,t>0)=0$, which translates to \mbox{$R_{1,k}(x,t)\propto \Theta (-t)$} ($\Theta$ being the Heaviside step function).

On the other hand, adding a regulator $R_{2,k}$ is equivalent to modifying the distribution of the noise, and, if we initially had a white noise, it means changing the noise correlations:
\begin{align}
    C(\bm x,\bm x') \propto\delta(\bm x-\bm x')  \, \rightarrow \,  \delta(\bm x-\bm x') - R_{2,k} (\bm x-\bm x')
    \label{eq:noise_correlations}
\end{align}
and the noise is now colored by the regulator along the flow and becomes $\delta$-correlated only at $k=0$ where $R_{2,k}$ must identically vanish.

Because we choose the \textit{ansatz}~(\ref{eq:derivativeExpansion}) to be invariant under the FDT transformation~(\ref{eq_invariance}), the regulator terms must also satisfy this symmetry along the flow. We show in Appendix \ref{sec_relationR1R2} that this implies that $R_{1,k}$ and $R_{2,k}$ satisfy the following relation:
\begin{align}
    R_{1,k} (\bm x) - R_{1,k} ( x,-t) + \dot R_{2,k} (\bm x) -\dot R_{2,k} (x,-t) = 0 \, .
\end{align}
The above condition, together with the facts that we choose $R_{1,k}$ to be causal and $R_{2,k}$  even in time (since it comes in $\int_{\bm x,\bm x'}\tilde\phi(\bm x) R_{2,k}(\bm x-\bm x')\tilde\phi(\bm x')$), lead to the following relation:
\begin{align}
R_{1,k}(\bm x)= 2  \Theta (- t) \, \dot R_{2,k}(\bm x) \, .
\label{eq:relation_R1R2}
\end{align}
Notice that the case $R_{2,k}(\bm x)\equiv 0$ which implies that $R_{1,k}(\vert x\vert,t)\propto \delta(t)$ is not included in the solutions of (\ref{eq:relation_R1R2}) which holds only for $t\neq 0$.
Eq.~(\ref{eq:relation_R1R2}) becomes in Fourier space:
\begin{align}
R_{2,k}(q,\omega)= \frac{R_{1,k}(q,-\omega)-R_{1,k}(q,\omega)}{2 i \omega}
\label{eq:relation_R1R2_Fourier}
\end{align}
where the Fourier transform is defined as (using abusively the same name for the function and its Fourier transform):
\begin{align}
    f(q,\omega) = \int_{\bm x} f(\bm x)\, \ee^{-i (q x - \omega t)}
    \label{eq:Fourier}
\end{align}
 Notice that the particular case $R_{2,k}\equiv 0$ and $R_{1,k}(q,\omega)$ independent of $\omega$ is a solution of (\ref{eq:relation_R1R2_Fourier}).

Specific choices of frequency regulators are given in Section~\ref{sect:reguexplicit}, and the results obtained using these regulators, and their comparison with the case without a frequency regulator, are shown in Section~\ref{sect:resultsregu}.

\section{Derivation of the RG equations}

In the previous sections we introduced the NPRG formalism in a formal way and we now give more details for the resolution of the flow equations. Since the formalism is the same for the multidimensional-spin counterpart of the model A, the kinetic $O(N)$ model, we focus in the following on the general case where the coarse-grained spin variable $\phi$ is  now a $N$-dimensional vector, denoted $\underline{\phi}$. We therefore modify the \textit{ansatz} for the effective average action $\Gamma_k$ to be
\begin{align}
    \begin{split}
  &\Gamma_k[\underline\psi,\underline\psit] =\int_{\bm x} \sum_i \tilde{\psi}_i \left[  X_k(\rho) \left( \partial_t \psi_i -\tilde{\psi}_i \right) + \psi_i U_k'(\rho) \right. \\
        &\left. \phantom{\tilde\psi}  +\frac{\psi_i}{2} Z_k'(\rho) (\nabla \underline\psi)^2 - Z_k(\rho) \nabla^2 \psi_i  - Z_k'(\rho)\nabla \psi_i (\underline\psi \cdot \nabla \underline \psi) \right. \\
    &\left. \phantom{\tilde\psi} +\frac{\psi_i}{4}Y_k'(\rho) (\nabla \rho)^2 + \frac{1}{2} Y_k(\rho) \nabla \rho \nabla \psi_i \right] \, ,
    \end{split}
\end{align}
where $\rho = \underline\psi^2/2$. In order to compute the RG flow of the functions involved in Eq.~(\ref{eq:derivativeExpansion}), we define them in the following way:
\begin{align}
    U'_k(\rho) &= \frac{1}{\psi} \left. \text{FT}\left( \left.  \frac{\delta \,\Gamma_k}{\delta \tilde\psi_1 (x)} \right|_{\scriptsize \underline\Psi=(\psi,\underline 0)} \right)  \right|_{\scriptsize \nu=0,p=0} \\
    Z_k(\rho) &= \left[ \partial_{p^2} \left. \text{FT}\left( \left. \frac{\delta^2\Gamma_k}{\delta \tilde\psi_2 (x) \delta \psi_2 (y)}\right|_{\scriptsize \underline\Psi=(\psi,\underline 0)} \right)   \right] \right|_{\scriptsize \nu=0,p=0}\\
    X_k(\rho) &= \left[ \partial_{i \nu} \left. \text{FT}\left( \left. \frac{\delta^2\Gamma_k}{\delta \tilde\psi_2 (x) \delta \psi_2 (y)}\right|_{\scriptsize \underline\Psi=(\psi,\underline 0)} \right)   \right] \right|_{\scriptsize \nu=0,p=0} 
\end{align}
where $\underline\Psi = (\psi,\underline0)$ is a $2N$ constant vector where $\Psi_1 = \psi_1 = \psi$ is a constant field and $\Psi_2 = \cdots = \Psi_{2N} = 0$, and $\text{FT}( \cdot )$ means the Fourier transform as defined in Eq.~(\ref{eq:Fourier}). Notice that in the case of the model A, the function $Y_k$ does not appear in the \textit{ansatz} for $\Gamma_k$, and that the functions $Z_k$ and $X_k$ are evaluated in the $\psi_1=\psi$, $\tilde\psi_1=\tilde\psi$ direction. In the spirit of the DE, we evaluate the renormalization functions at zero external momentum and frequency since it is in this limit that the approximation is valid. The flow of these functions is then computed using the Wetterich equation~(\ref{eq:Wetterich}) with the initial conditions $U'_\Lambda=V'$, $Z_\Lambda=1=X_\Lambda$. 

As an example, the flow of $U'_k$ for the model A (i.e. a one-dimensional spin) is given by
\begin{align}
    \partial_k U'_k(\rho) &= \frac{1}{\psi} \text{FT}\left( \left.  \frac{\delta }{\delta \tilde\psi} \partial_k \Gamma_k \right|_{\scriptsize \Psi=(\psi,0)} \right) \\
    &= \frac{1}{\psi} \text{FT}\left( \frac{1}{2} \tilde\partial_k \text{Tr}  \left.  \left[ \int_{t_i,x_i}  \Gamma_{k,\tilde\psi}^{(3)} \cdot G_k  \right]  \right|_{\scriptsize \Psi=(\psi,0)} \right)
    \label{eq:flowUp}
\end{align}
where $\Gamma_{k,\tilde\psi}^{(3)} \equiv \delta \Gamma^{(2)}_k/\delta \tilde\psi$ and $\tilde\partial_k \equiv \partial_k R_k \, \partial/\partial R_k$. Taking the appropriate functional derivatives of (\ref{eq:derivativeExpansion}) and evaluating the result at the uniform and stationary field configuration $\Psi (x,t) = (\psi,0)$, one finds in Fourier space:
\begin{align}
    & \Gamma^{(3)}_{k,\tilde\psi} (p,\nu; q,\omega; \psi) = \psi
    \left( 
         \begin{array}{cc} h_3(p,\nu; q,\omega; \rho) & -2 X'_k(\rho) \\ -2 X'_k(\rho) &  0 \end{array} 
    \right)
\end{align}
with $h_3=2\rho \, U^{(3)}_k+3U''_k+Z'_k(p^2+q^2+p\cdot q) - i \nu X'_k$, which is a function of $\rho$. The propagator $G_k$ in Eq.~(\ref{eq:flowUp}) is obtained by inverting the $2\times 2$ matrix $(\Gamma_k^{(2)} + \mathcal{R}_k)$ evaluated at $\Psi (x,t) = (\psi,0)$. One finds:
% \begin{align}
% \begin{split}
%     &G_k(q,\omega;\rho) = \\
%     &\left( 
%          \begin{array}{cc} 
%                 \frac{-R_{2,k}(q^2,\omega)+2X_k}{(h(q^2,\omega)+i \omega X_k)(h(q^2,-\omega)-i \omega X_k)} & \frac{1}{h(q^2,-\omega)-i \omega X_k} \\ 
%                 \frac{1}{h(q^2,\omega)+i \omega X_k} &  0 
%          \end{array} 
%     \right) 
%     \end{split}
% \end{align}
\begin{align}
    G_k(q,\omega;\rho) = 
    \left( 
         \begin{array}{cc} 
                \frac{-R_{2,k}(q^2,\omega)+2X_k}{P(q^2,\omega)P(q^2,-\omega)} & \frac{1}{P(q^2,-\omega)} \\ 
                \frac{1}{P(q^2,\omega)} &  0 
         \end{array} 
    \right) 
\end{align}
where $P(q^2,\omega)=h(q^2,\omega)+i \omega X_k$ with $h(q^2,\omega) = Z_k(\rho)q^2+R_{1,k}(q^2,\omega)+U'_k(\rho)+2\rho U''_k(\rho)$.

\subsection{Definition of the dimensionless variables and functions}

Since we are interested in the scale invariant regime, we introduce the dimensionless and renormalized variables, fields and  functions:
\begin{subequations}
\begin{empheq}{align}
    \hat{x}&=k \,x\\
    \hat t&=\bar{Z}_k\bar{X}_k^{-1}
    k^{2}\,t\\
    \hat{\underline\psit}(\hat x,\hat t\,) &= k^{(2-d)/2} \bar{Z}_k^{1/2} \underline\psit(x,t) \\
    \hat{\underline\psi}(\hat x,\hat t\,) &= k^{(2-d)/2} \bar{Z}_k^{1/2} \underline\psi(x,t) \\
    \hat\rho(\hat x,\hat t\,) &= k^{2-d} \bar{Z}_k \rho(x,t) \\
    \hat{U}(\hat\rho) &= k^{-d} U_k(\rho) \\
    \hat{Z}(\hat\rho) &= \bar{Z}_k^{-1}Z_k(\rho) \\
    \hat{X}(\hat\rho) &= \bar{X}_k^{-1}X_k(\rho)
\end{empheq}
\end{subequations}
where the running coefficients $\bar{Z}_k \equiv Z_k(\rho_0)$ and $\bar{X}_k \equiv X_k(\rho_0)$ are defined at a fixed normalization point $\rho_0$. In the critical regime, these running coefficients are expected to behave as power laws $\bar{Z}_k \sim k^{-\eta(k)}$ and $\bar{X}_k \sim k^{-\eta_X(k)}$ with $\eta(k) = - k \partial_k \ln \bar Z_k$ and similarly for $\eta_X(k)$. The anomalous dimension $\eta$ and the dynamical exponent $z$ can be expressed in terms  of the fixed point values of $\eta(k)$ and $\eta_X(k)$ as $\eta \equiv \eta^*$ and $z\equiv 2 -\eta^* +\eta_X^*$.

We furthermore define the dimensionless regulators $r_1$ and $r_2$ such that :
% \begin{align}
%     R_{1,k}(q,\omega) &= q^2 \bar{Z}_k r_1 (y, \hat{\omega}) \\
%     &= y \bar{Z}_k k^2 \rho_1(\hat{\omega}) r(y) \\
%     R_{2,k}(q,\omega) &= \bar{X}_k r_2 (y, \hat{\omega}) \\
%     &= y \bar{X}_k \rho_2(\hat{\omega}) r(y)
% \end{align}
\begin{align}
    R_{1,k}(q,\omega) &= y \bar{Z}_k k^2 r_1 (y, \hat{\omega}) \nonumber\\
    &= y \bar{Z}_k k^2 \rho_1(\hat{\omega}) r(y) \\
    R_{2,k}(q,\omega) &= \bar{X}_k r_2 (y, \hat{\omega}) \nonumber\\
    % &= w(y)\bar{X}_k \rho_2(\hat{\omega}) r(y)
    &= y \bar{X}_k \rho_2(\hat{\omega}) r(y)
\end{align}
with $y=\hat q^2$ and $\hat{\omega}=\bar{X}_k\bar{Z}_k^{-1}
    k^{-2}\omega$ and
where we have assumed for simplicity that the spatial and frequency parts of the regulators can be factorized, and where $r(y)$ is the usual momentum regulator, for example an exponential regulator:
\begin{align}
r(y) =  \frac{\alpha}{\ee^y-1}
\label{eq:reguExp}
\end{align}
where $\alpha$ is a free parameter. The frequency part of the regulators, $\rho_1$ and $\rho_2$, have to satisfy condition (\ref{eq:relation_R1R2_Fourier}), and we give explicit examples in the following.

Notice already that from the previous definitions we deduce the regulator derivatives with respect to $k$:
\begin{align}
\begin{split}
    k \partial_k R_{1,k}(q,\omega) = - k^2 & \bar{Z}_k  y (\eta r_1 + 2 y \partial_y r_1 \\
    &+(2-\eta+\eta_X) \hat{\omega} \partial_{\hat{\omega}} r_1 ) 
\end{split} \\
\begin{split}
     k \partial_k R_{2,k}(q,\omega) =  -\bar{X}_k & (\eta_X r_2 + 2 y \partial_y r_2 \\
     &+(2-\eta+\eta_X) \hat{\omega} \partial_{\hat{\omega}} r_2 )  \, .
\end{split}
\end{align}

Finally, using dimensionless variables yields a part for the flow equations which is purely dimensional. Thus, the dimensional parts for the flows of $\hat{U}'$, $\hat{Z}$ and $\hat{X}$ are respectively:
\begin{align}
    k\partial_k \hat{U}' |_{\text{dim}} &= (\eta - 2)\,\hat{U}' + (d+\eta-2)\, \hat\rho \, \hat{U}''
    \label{eq:dimPartu} \\
    k\partial_k \hat{Z} |_{\text{dim}} &= \eta \, \hat{Z} + (d+\eta-2) \hat\rho \, \hat{Z}'
    \label{eq:dimPartz} \\
    k\partial_k \hat{X} |_{\text{dim}} &= \eta_X \, \hat{X} + (d+\eta-2) \hat\rho \, \hat{X}'
    \label{eq:dimPartx}
\end{align}

\begin{figure}[t!]
	\centering 
    \subfigure[ ]
    {\includegraphics[width=0.42\textwidth]{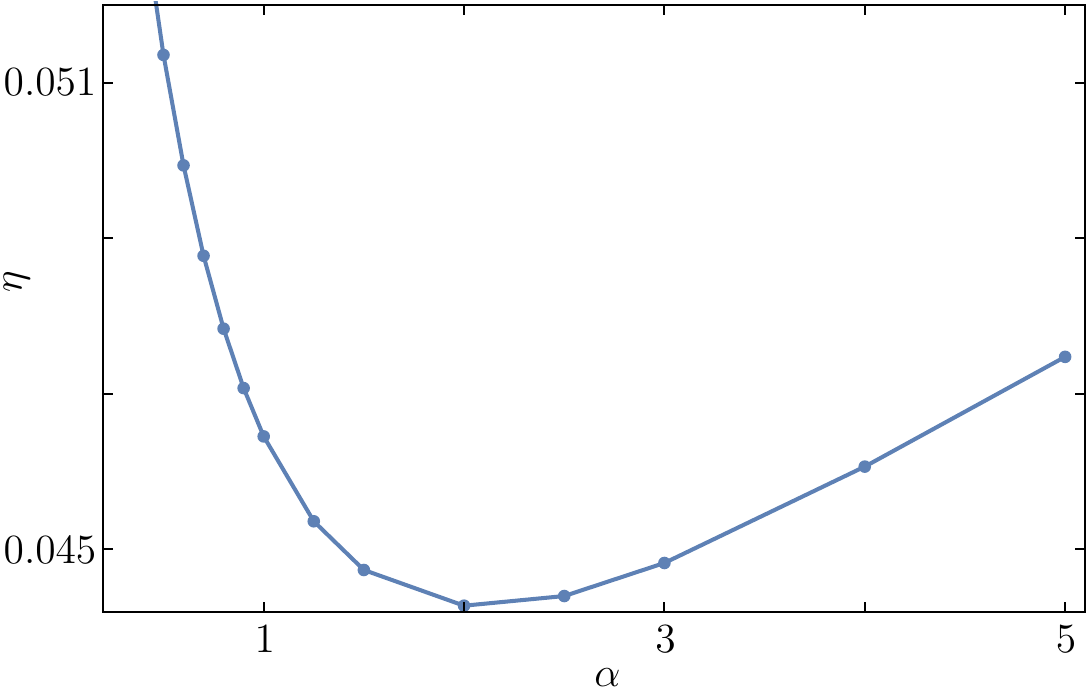}}
    \subfigure[ ]
    {\includegraphics[width=0.42\textwidth]{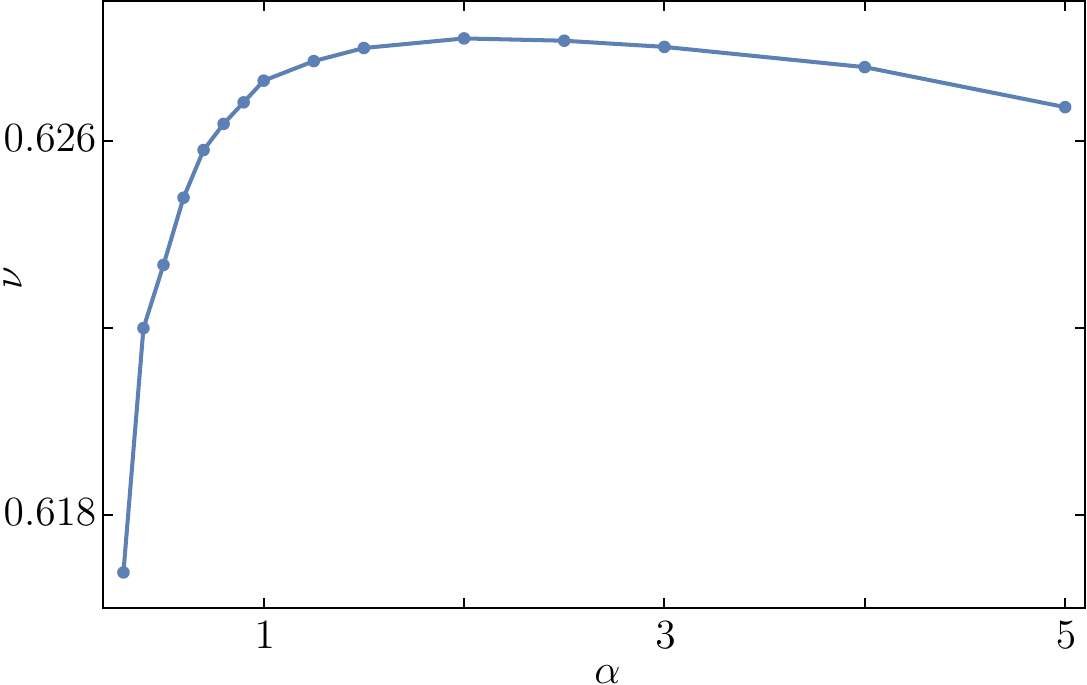}}
    \subfigure[ ]
    {\includegraphics[width=0.42\textwidth]{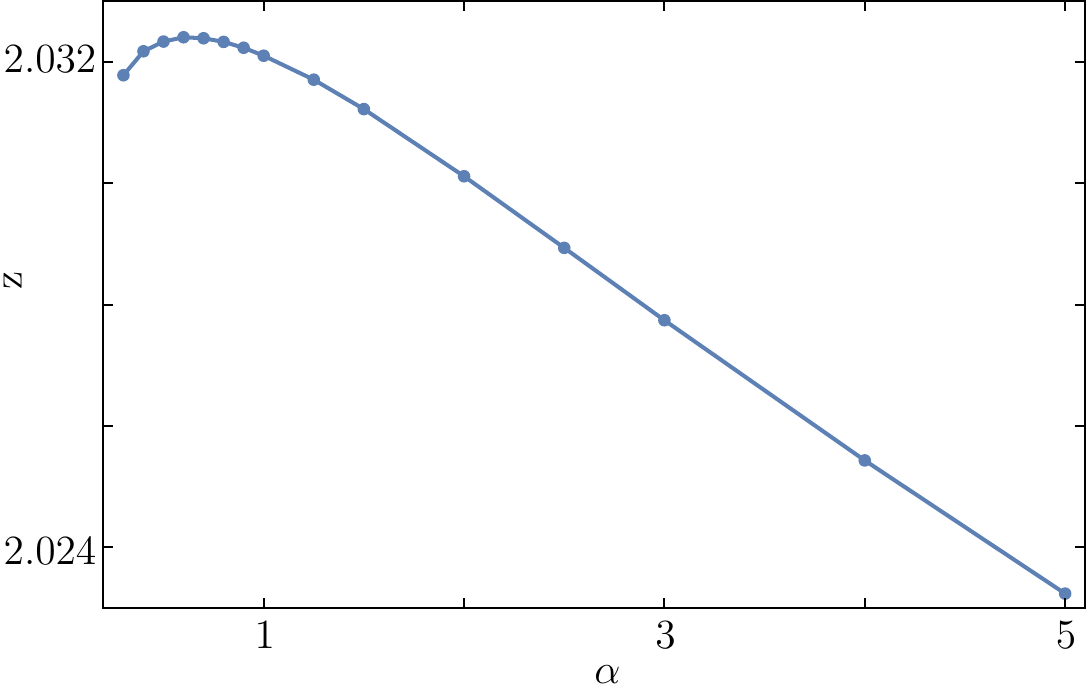}}

    \caption{Values of the critical exponents $\eta$ (a), $\nu$ (b) and the dynamical exponent $z=2+\eta_X-\eta$ (c) in $d=3$ for the frequency-independent regulator $ R_{1,k}(q,\omega)= R_{1,k}(q), \ R_{2,k}(q,\omega)=0 $ and different values of the regulator parameter $\alpha$ in Eq.(\ref{eq:reguExp}). The PMS value is reached for $\alpha \simeq 2$ for the two static exponents $\eta$ and $\nu$ and for $\alpha\simeq0.6$ for the dynamical exponent $z$.}
    \label{fig:pms_model_A}
\end{figure}

  \section{Results}
  \label{sect:flow-eqs}

\subsection{Results without a frequency regulator}
In a first step, we consider frequency-independent regulators, which means $R_{2,k}=0$ and $R_{1,k}(q,\omega)=R_{1,k}(q)$. In this case, the calculation of the flow equations is much simpler since the integration over frequency can be done analytically using residues, see Appendix~\ref{sec_ONflow} for the explicit expression of the flow equations in this case.

Contrary to the Ising case where we keep the full $\rho$-dependence of the functions $U_k, Z_k$ and $X_k$,  we perform in the $O(N)$ case, on top of the DE, a field expansion usually called the Local Potential Approximation prime (LPA'). It consists in discarding the function $Y_k(\rho)$ and neglecting the field dependence of  $Z_k(\rho)$ and $X_k(\rho)$:  $Z_k(\rho)\to \bar{Z}_k$ and $X_k(\rho)\to \bar{X}_k$.
The dynamical part of their flow equation is given in  Appendix~\ref{sec_ONflow}.

Notice that the flows of $\hat{U}'(\hat\rho)$ and $\hat{Z}(\hat\rho)$ do not depend on $\hat{X}(\hat\rho)$ and are the standard equilibrium flow equations of the Ising model: This is not surprising because  with the regulators chosen above, the model A satisfies for any $k$ the FDT  which is the hallmark of thermal equilibrium. Consequently, the critical exponents $\nu$ and $\eta$ for the model A are the same as in the static Ising model. 

Our results are optimized with respect to the parameter $\alpha$ of the regulator using the Principle of Minimum Sensitivity (PMS) \cite{Stevenson81,Canet03,Canet03_2}: In the exact theory, the critical exponents do not depend on the unphysical parameter $\alpha$ and we therefore select the values of this parameter where the exponents are stationary; see Fig.~(\ref{fig:pms_model_A}) for model A in $d=3$. 

The numerical integration of the  flow equations~(\ref{eq:flowup})-(\ref{eq:flowx}) for the model A yields the results displayed in Table~\ref{tableSSregud3} for $d=3$ together with the results coming from  Perturbative Field Theory (PFT), Monte-Carlo (MC) simulations and previous NPRG works where the field-dependence of the functions $Z_k$ and $X_k$ was neglected.
For $d=2$, the results are given in Table~\ref{tableSSregud2}.

Similarly, for $N=2$ and $N=3$, the integration of the equations~(\ref{eq:dynON1})-(\ref{eq:dynON3}) yields the results displayed in Table~\ref{tableSSreguON}. Note that  an expansion of these equations in $\epsilon'=d-2$ yields $\eta=\eta_X=\epsilon'/(N-2)$ and therefore a trivial dynamical exponent $z=2$ in $d=2$ for $N>2$.

Finally, notice in the plots of Fig.~\ref{fig:pms_model_A} that stationarity  yields  values of $\alpha$ that are close to each other for both  $\eta$ and $\nu$: $\alpha_{\eta,{\rm PMS}}\simeq\alpha_{\nu,{\rm PMS}}\simeq2$, whereas the PMS for $z$ is obtained when $\alpha_{z,{\rm PMS}}\simeq0.6$. The internal consistency of the PMS relies on the fact that the values of an exponent computed either at its stationary point or at the stationary points of the other exponents remain close. This is not the case here since we find for instance that $\eta(\alpha=\alpha_{z,{\rm PMS}})=0.0499$ which differs by about 13$\%$ from its PMS value whereas $\eta(\alpha=\alpha_{\nu,{\rm PMS}})$ and $\nu(\alpha=\alpha_{\eta,{\rm PMS}})$ differ from their PMS values by less than 1$\%$. This is a signal that the exponent $z$ is poorly determined and it is therefore mandatory to study the impact of the frequency-dependence of the regulator on  this exponent.

\begin{table}[t]
\centering
    \begin{ruledtabular}
        \begin{tabular}{lllll}
        Reference & $\nu$ & $\eta$ & $z$ &  \\
        \hline
        This work & 0.628 & 0.0443 & (a): 2.032 (b): 2.024 &  \\
        &   &  &  (c): 2.024 (d): 2.023 & \\
        &   &  &   & \\
        NPRG  & 0.6281 \cite{Canet05} & 0.0443 \cite{Canet05} & 2.14 \cite{Canet07} & \\
        PFT \cite{Guida98} & 0.6304(13) & 0.0335(25) & & \\
        MC \cite{Hasenbusch10} & 0.63002(10) & 0.03627(10) & &  \\
        CBS \cite{Kos16} & 0.629971(4) & 0.036298(2) & &  \\
        PFT \cite{Krinitsyn06} & & & 2.0237(55) & \\
        MC \cite{Ito00} & & & 2.032(4) & \\
        \end{tabular}
    \end{ruledtabular}
    \caption{Critical exponents of model A in $d=3$ from different methods. In the first row, (a): without frequency regulator, (b): using the first frequency regulator defined by Eq.~(\ref{eq:R1}), (c): second regulator~(\ref{eq:regu2}) and (d): third regulator~(\ref{eq:regu3}). All these results are obtained at the stationary points, $\alpha_{\rm PMS}$. The exponent $z$ in the NPRG row was obtained in \cite{Canet07}, where the field-dependence of the functions $Z_k$ and $X_k$ was neglected. PFT stands for Perturbative Field Theory methods, MC for Monte Carlo simulations, and CBS for Conformal Bootstrap methods.}
\label{tableSSregud3}
\end{table}

\begin{table}[t]
\centering
    \begin{ruledtabular}
        \begin{tabular}{lllll}
        Reference & $\nu$ & $\eta$ & $z$ &  \\
        \hline
        This work & 1.13 & 0.29 & (a): 2.28  &  \\
        &   &  & (b): 2.16 (c): 2.15 (d): 2.14  & \\
        Exact  & 1 & 0.25 &  & \\
        PFT \cite{Prudnikov97} &  &  & 2.093 &\\
        MC \cite{Nightingale00} &  &  & 2.1667(5) & 
        \end{tabular}
    \end{ruledtabular}
    \caption{Critical exponents of model A in $d=2$ from different methods. In the first row, (a): without frequency regulator, (b): using the frequency regulator defined by Eq.~(\ref{eq:R1}), (c): second regulator~(\ref{eq:regu2}) and (d): third regulator~(\ref{eq:regu3}). All these results are obtained at the stationary points, $\alpha_{\rm PMS}$. We also display results for the dynamical exponent $z$ coming from Perturbative Field Theory (PFT) and Monte Carlo (MC) simulations.}
\label{tableSSregud2}
\end{table}

\begin{table}[t]
\centering
    \begin{ruledtabular}
        \begin{tabular}{lllll}
        Reference & $\nu$ & $\eta$ & $z$ &  \\
        \hline
        This work ($N=2$) & 0.70 & 0.039 & (a): 2.029 &  \\
        &   &  &  (b): 2.024 (c): 2.023 & \\
        This work ($N=3$) & 0.75 & 0.037 & (a): 2.025 &  \\
        &   &  &  (b): 2.021 (c): 2.021 & \\
        &   &  &   & \\
        PFT ($N=2$)  & 0.6704(7) & 0.0349(8) & 2.026 &\\
        PFT ($N=3$)  & 0.7062(7) & 0.0350(8) & 2.026 & 
        \end{tabular}
    \end{ruledtabular}
    \caption{Critical exponents of the kinetic $O(N)$ model in $d=3$ for different values of $N$ and from different methods. The exponents $\eta$ and $z$ have been computed in this work using the LPA' (see section \ref{sect:flow-eqs} in the main text for a definition). In the two first row, (a): without frequency regulator, (b): using the frequency regulator defined by Eq.~(\ref{eq:R1}) and (c): second regulator~(\ref{eq:regu2}). All these results are obtained at the stationary points, $\alpha_{\rm PMS}$. The static exponents $\eta$ and $\nu$ for the Perturbative Field Theory (PFT) comes from \cite{Jasch01}, the dynamic exponent $z$ is computed using the value of $\eta$ from \cite{Jasch01} and the relation $z=2+c\eta$ from \cite{Adzhemyan08}, which is a relation obtained perturbatively  at order $\epsilon^4$, with $\epsilon=4-d$. Very few MC studies exist for the determination of $z$. More details on the determination of this exponent can be found in \cite{Calabrese05} and see also reference \cite{Pelissetto02} for a review of the determination of the static exponents.}
\label{tableSSreguON}
\end{table}

    \subsection{Specific choices of frequency regulators}
    \label{sect:reguexplicit}

\begin{figure}[t!]
	\centering 
    {\includegraphics[width=0.45\textwidth]{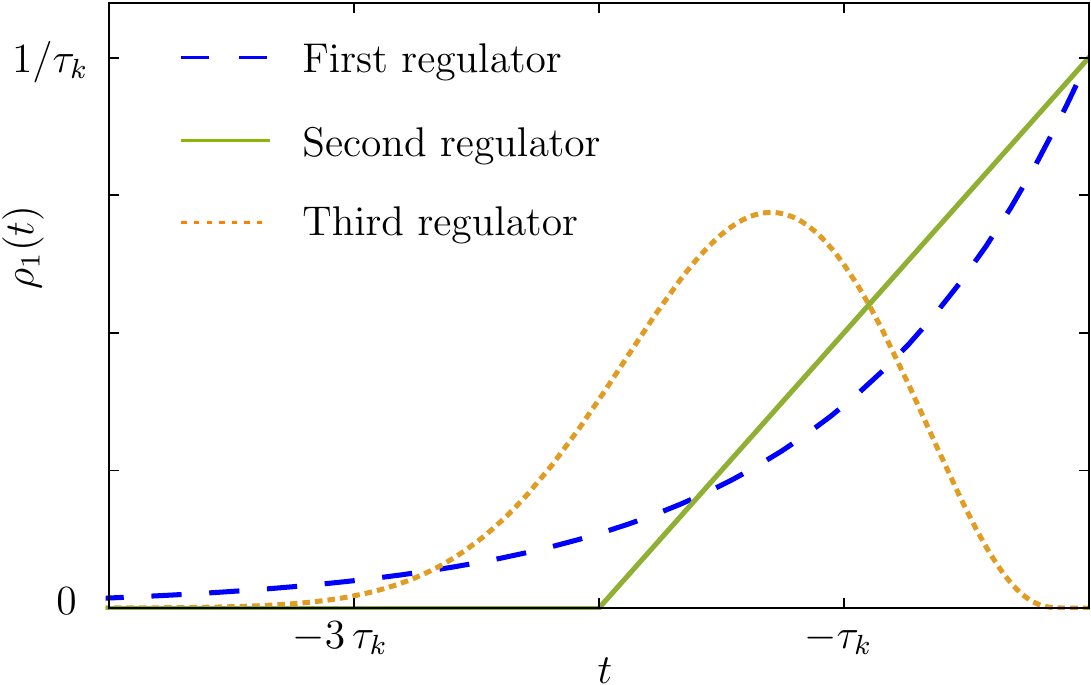}}
    \caption{Typical shape of the time-dependent part $\rho_1(t)$ of the three regulators studied: the first regulator is defined in Eq.(\ref{eq:R1}), the second in (\ref{eq:regu2}) and the third in (\ref{eq:regu3}).}
    \label{fig:definition_regus}
\end{figure}

We now focus on regulating the frequencies in the flow equations. We have discussed the general constraints a frequency regulator should fulfill in Section~\ref{sect:regu}, and we present here the three different regulators -- all suited for regulating large frequencies but not equally efficient -- we will use for computing $z$.

A convenient choice for the regulator in direct space is the following:
\begin{align}
   R_{1,k} (x,t) &=  \frac{1}{\tau_k}  \Theta (-t) \, \mathrm{e}^{t/\tau_k} r_k(x) \label{eq:R1} %\\
%   R_{2,k} (x,t) &= \frac1 2 \, \mathrm{e}^{-|t|/\tau_k} r_k(x) \label{eq:R2}
\end{align}
where $r_k(x)$ is the space regulator (usually exponential) whose Fourier transform is given by Eq.~(\ref{eq:reguExp}), and $\tau_k=\beta \, \bar{X}_k \bar{Z}_k^{-1} k^{-2}$ with $\beta$ a dimensionless free parameter that we use for optimization. We display the time-dependent part $\rho_1(t)$ of this regulator in Fig.~\ref{fig:definition_regus}. The choice of this first regulator is motivated by three main reasons: (i) it is causal and satisfies relation~(\ref{eq:relation_R1R2}), (ii) it decays sufficiently fast in time so that the noise correlations~(\ref{eq:noise_correlations}) are not modified too drastically, (iii) its Fourier transform can be computed analytically and is a simple rational fraction. Indeed, using dimensionless frequencies, the Fourier transforms of their frequency part reads:
\begin{align}
   \rho_1(\hat{\omega}) &= \frac{i}{i-\beta \hat{\omega}} \, ,\\
   \rho_2(\hat{\omega}) &= \frac{\beta}{1+\beta^2 \hat{\omega}^2} \, .
\end{align}
When $\beta \to 0$, we retrieve the usual non-regulated in time theory. Now that we have specified the frequency and space parts of the regulators, we check that they both fulfill NPRG requirements: in addition to a sufficiently fast decay, they should also satisfy some limits when $k\to 0$ and $k\to \Lambda$: $R_{1,k}(q,\omega)$ and $R_{2,k}(q,\omega)$ should both vanish when $k\to0$ in order to retrieve the original theory. In the limit $k\to \Lambda$, we design $R_{1,k}$ such that $R_{1,k}(q,\omega)\subrel{k\to\Lambda}{\sim}\Lambda^2 \gg 1$: the system acquires a large ``mass'' that freezes the fluctuations. Finally, one finds $R_{2,k}(q,\omega) \subrel{k\to\Lambda}{\sim} \alpha \beta$, which means the initial noise correlation is modified which is harmless for the computation of universal quantities.

In order to compare the results obtained with different frequency regulators, we have engineered two other regulators in addition to this simple first one, see Fig.~\ref{fig:definition_regus} for a plot of their time-dependent part. 

The second regulator we propose is defined as:
\begin{align}
      R_{1,k}(x,t) = \frac{r_k(x)}{2\tau_k} \times
     \left\{
     \begin{array}{rl}
     (t+2\tau_k)/\tau_k & \text{if } -2\tau_k \leq t \leq 0,\\
     0 & \text{otherwise}.
     \end{array}
     \right.
     \label{eq:regu2}
\end{align}
 Notice that its Fourier transform can also be computed analytically. 
Since singularity in the time domain means slow decay in the frequency domain, the more singular in $t$ the slower the decay of $\rho_1(\hat\omega)$ at large $\hat\omega$. This second regulator has two singularities (at $t=0$ and $t=-\tau_k$) and we therefore expect it to be less effective than the first one.

Finally, the third frequency regulator we consider is the following:
\begin{align}
R_{1,k}(x,t) = \frac{A}{\tau_k} \Theta (-t) \, \ee ^{-(1+t/\tau_k)^2 + \tau_k/t} r_k(x)
\label{eq:regu3}
\end{align}
where $A$ is a constant such that the area under its curve is one, in order to retrieve a Dirac function as $\beta \to 0$. This third regulator is infinitely differentiable at $t=0$ and we therefore expect it to be sharper than the two previous regulators in the frequency domain. On the other hand, the computation of its Fourier transform has to be done numerically. 

Finally, we insist on the fact that enforcing causality along the flow is not an obvious task: Although choosing a regulator that is causal ($R_{1,k}(x,t) \propto \Theta (-t)$) seems at least necessary to preserve causality, one should check that it also preserves causality all along the flow~\cite{Canet11}. As we explain in Appendix~\ref{sect:KK}, causality means that the poles of the response function 
\begin{align}
    \chi (\omega) &= \frac{1}{P(q^2,-\omega)} = \frac{1}{h(q^2,-\omega)-i \omega X_k} \, ,
\end{align}
where $h(q^2,\omega) = Z_k(\rho)q^2+R_{1,k}(q^2,\omega)+U'_k(\rho)+2\rho U''_k(\rho)$, must lie in the lower-half of the complex $\omega$-plane. When $R_{1,k}(q^2,\omega)$ is a (simple) rational fraction as it is the case for the first regulator defined by Eq.~(\ref{eq:R1}),it is easy to check that the causality of the response function is enforced all along the flow. For the second regulator~(\ref{eq:regu2}) and the third regulator~(\ref{eq:regu3}), we only checked it for the initial condition, and at the fixed point of the flow. 

We also stress that if $R_{1,k}(q^2,\omega)$ is a rational fraction, one can hope to design ``by hand'' a regulator for which all the poles of the response function have a negative imaginary part. However, if one wishes to build a regulator that decays faster than a power-law, then the only remaining option is to construct it in direct space and afterwards check the decay in Fourier space.

\subsection{Results with a frequency regulator}
\label{sect:resultsregu}

\begin{figure}[t!]
	\centering 
    %\subfigure[ ]
    {\includegraphics[width=0.45\textwidth]{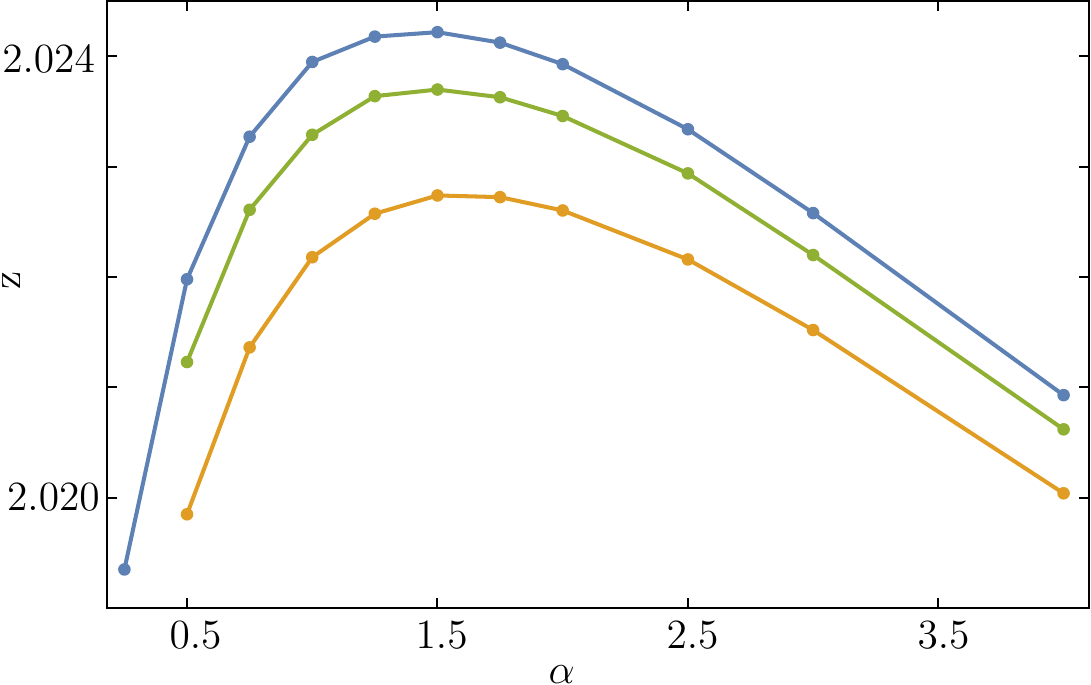}}
    \caption{Color online. Values of the critical  exponent $z$ in $d=3$ for the flow regulated in frequencies for  different values of the regulator parameter $\alpha$. For each value of $\alpha$, the value of $\beta$ has been chosen such that $z$ is extremal. The three curves are obtained from top to bottom by the regulators defined in Eqs.~(\ref{eq:R1}), (\ref{eq:regu2}) and (\ref{eq:regu3}). The PMS value is reached for $\alpha \simeq 1.5$ for the three regulators.}
    \label{fig:PMSregufreq}
\end{figure}

\begin{figure}[t!]
	\centering 
    {\includegraphics[width=0.45\textwidth]{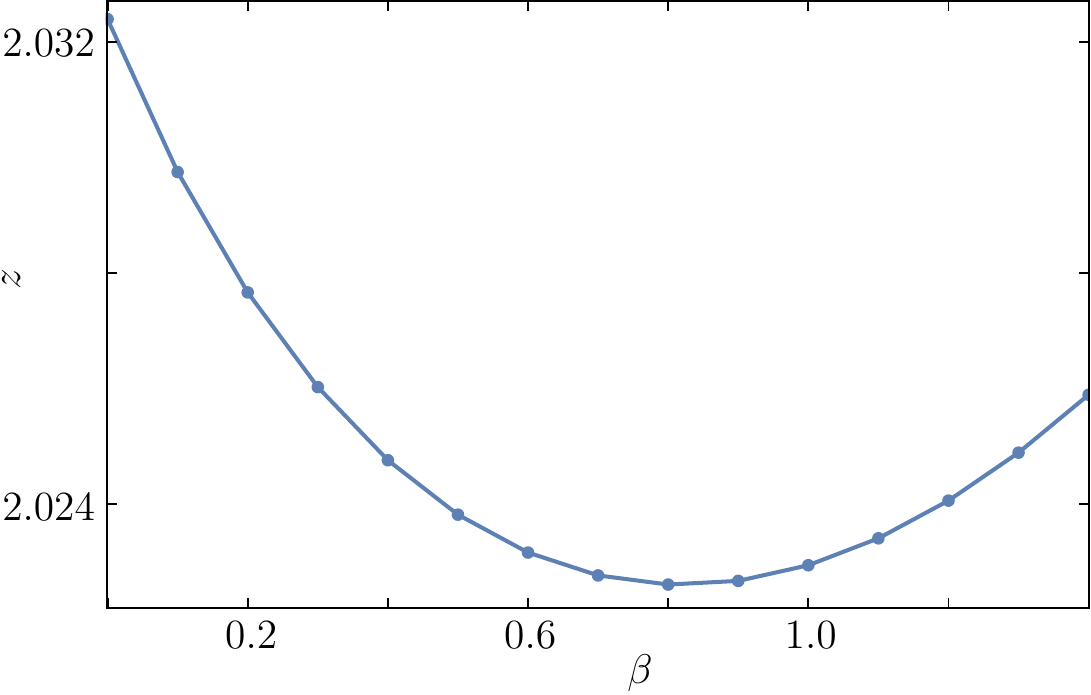}}
    \caption{Exponent $z$ for $N=1$ in $d=3$ as function of the parameter $\beta$ of the regulator~(\ref{eq:R1}). This curve is obtained for $\alpha=0.6$ which corresponds to the stationary point of $z$ at $\beta=0$.}
    \label{fig:alpha06}
\end{figure}

In the presence of the three regulators defined respectively in Eqs.~(\ref{eq:R1},\ref{eq:regu2},\ref{eq:regu3}), the flow equations of 
$\hat{U}'$ and $\hat{Z}$ remain identical to those at equilibrium~(\ref{eq:flowup})-(\ref{eq:flowz}) since the FDT is valid all along the flow. On the other hand, the flow of $\hat{X}$ now depends on $\beta$ and is more complicated than without a frequency regulator. For the first regulator defined by Eq.~(\ref{eq:R1}), the integrals over the frequencies in the flow equation can still be performed analytically since its Fourier transform is a simple rational fraction in $\hat{\omega}$. For the two other regulators, the integrals over frequencies must be computed numerically.

We have numerically integrated the new flow equations for different values of $\alpha$ and $\beta$ in order to compute the critical exponents at the stationary point in the $(\alpha,\beta)$-plane. For each value of  $\beta$, we find a value of $\alpha$ where $z$ is extremal. This yields a curve  $z(\alpha)$, see Fig.~(\ref{fig:PMSregufreq}), that shows a maximum which is therefore the stationary point in the $(\alpha,\beta)$-plane. One notices that the PMS value is now obtained for $\alpha_{z,{\rm PMS}} \simeq 1.5$ (instead of $0.6$ in the case without a frequency regulator), which is closer to the PMS value of $\eta$ and $\nu$ (obtained at $\alpha\simeq 2$). More precisely, we find for instance for the model A in $d=3$ that $\eta(\alpha=\alpha_{z,{\rm PMS}})$  differs by about 1$\%$ from its PMS value, and $\nu(\alpha=\alpha_{z,{\rm PMS}})$ and $z(\alpha=\alpha_{\nu,{\rm PMS}})$ differ from their PMS values by less than 1$\%$.

In the light of the above, it is clear that the frequency-independent regulators are simply a particular class of regulators. In our examples, they correspond to the limit $\beta\to 0$ of the three frequency regulators studied.  Their  main advantage is their simplicity since there is only one regulator which lies in the $\tilde\phi-\phi$ direction and also because the frequency integrals can be performed analytically in the flow equations. However, we can see in Fig.~\ref{fig:alpha06} that from the point of view of the PMS, the class of regulators with $\beta=0$ does not correspond at all to an extremum in the $\beta$-direction, even for the value $\alpha=0.6$, which is optimal at $\beta=0$. Moreover, the difference between $\alpha_{z,{\rm PMS}} \simeq 1.5$ and $\alpha=0.6$   is not only quantitatively important, it is also qualitatively important because it makes the PMS a self-consistent criterion for optimizing the critical exponents. It is remarkable and reassuring that this latter value of $\alpha_{z,{\rm PMS}}$, which has a meaning {\it per se} because it can be compared to  $\alpha_{\eta,{\rm PMS}}\simeq\alpha_{\nu,{\rm PMS}}\simeq2$, is extremely stable when changing the shape of the regulator, see Figs. \ref{fig:definition_regus} and \ref{fig:PMSregufreq}. Finally, we find as expected that the accuracy of the optimized value of $z-2$ found in this work  compared to the average of the other estimates, $z-2\simeq 0.028$, is comparable to the accuracy of the optimized value of $\eta$ compared to the world's best value, that is, is around $15\%$, see Table.~\ref{tableSSregud3}. Together with the stability of our results, this is a strong indication that the regulators we study here are almost optimal at this order of the DE.

\section{Conclusion}

We have engineered in this article regulators of the NPRG flow equations acting on frequencies, a feature that we believe can be of tremendous importance if we aim to solve generic out-of-equilibrium problems with the derivative expansion. Causality, of course, has to be taken care of and is the main preoccupation when designing such a regulator. Therefore, contrarily to the space regulator which can be engineered directly in  Fourier space, it is convenient to think first in direct space for a frequency regulator to enforce causality. For systems that relax towards equilibrium, introducing a second regulator in the $\tilde\phi-\tilde\phi$ direction connected to the other one in the $\tilde\phi-\phi$ direction is mandatory to preserve the time-reversal symmetry all along the flow, a feature that is surely desirable and that, at least, simplifies the formalism. 

The next step will be to implement frequency regulators for generic out of equilibrium  models not displaying such a strong constraint as the FDT. In the previous NPRG studies of the Directed Percolation universality class, only results at the LPA' were reported~\cite{Canet04_2,Canet06}.
Improving these results by going at order two of the DE surely requires the use of a frequency regulator. The Parity Conserving Generalized Voter model is another candidate since the NPRG results are not fully satisfactory for this model; see \cite{Benitez13} for  an exact result that disagrees with the conclusions of~\cite{Canet05_2} obtained within the LPA without frequency regulators.

\section*{Acknowledgements}

We thank Nicol\'as Wschebor, Gilles Tarjus and Matthieu Tissier for useful discussions.

\appendix
\section{Relation between $R_1$ and $R_2$}
\label{sec_relationR1R2}

We show here that the invariance of the action under transformation~(\ref{eq_invariance}) enforces constraints on $R_1$ and $R_2$. Let us define
\begin{align}
     \Delta \mathcal{S}_1&=\int_{\bm x,\bm{x'}} \tilde{\phi}(x,t) R_1(x'-x,t'-t) \phi (x',t') \, ,\\
     \Delta \mathcal{S}_2&=\int_{\bm x,\bm{x'}} \tilde{\phi}(x,t) R_2(x'-x,t'-t') \tilde{\phi}(x',t') \, ,
\end{align}
in which, for notational convenience, we drop in the following the spatial dependence in the different terms. After transforming the fields by (\ref{eq_invariance}), $\Delta \mathcal{S}_i[\phi,\tilde\phi]$ become $\Delta \mathcal{S}_i[\phi',\tilde\phi'] \equiv \Delta \mathcal{S}_i'[\phi,\tilde\phi]$ which read:
\begin{align}
     \Delta \mathcal{S}_1' &=\int_{t,t'} \tilde{\phi}(t) R_1(t-t') \phi (t') - \int_{t,t'} \dot \phi(t) R_1(t'-t) \phi (t') \, ,     \\
\begin{split}
     \Delta \mathcal{S}_2' &=\int_{t,t'} \tilde{\phi}(t) R_2(-t'+t)\tilde{\phi}(t') -\int_{t,t'} \tilde{\phi}(t) \left[ R_2(-t'+t) \right. \\
     & \left. + R_2(t'-t) \right]  \dot \phi(t')
     +\int_{t,t'}  \dot \phi(t) R_2(-t'+t)  \dot \phi(t') \, .  
\end{split}
\end{align}
In $\Delta \mathcal{S}_2'$ we notice that the first term gives back $\Delta \mathcal{S}_2$, and the third term is symmetric in $t$ and $t'$ and can be rewritten as:
\begin{align}
    \frac{1}{2} \int_{t,t'}  \dot \phi(t) \left( R_2(-t'+t) + R_2(t'-t) \right)  \dot \phi(t')
\end{align}
The invariance of the action under transformation~(\ref{eq_invariance}) yields the equality $\Delta \mathcal{S}_1'+\Delta \mathcal{S}_2'=\Delta \mathcal{S}_1+\Delta \mathcal{S}_2$, that reads:
\begin{align}
\begin{split}
    &\int_{t,t'}  \tilde{\phi}(t)  \left( R_1(-t'+t) + \dot R_2(-t'+t) \right. \\
    & \left. - \dot R_2(t'-t) -R_1(t'-t) \right) \phi (t') + \\
    &\int_{t,t'} \dot \phi (t) \left( -R_1(-t'+t) \phantom{\frac 1 2} \right. \\
    & \left. + \frac{1}{2}\left(\dot R_2(-t'+t) - \dot R_2(t'-t) \right) \right) \phi (t')    =0 \, .
\end{split}
\label{eq-R-temp}
\end{align}
which should be valid for all fields $\phi$ and $\tilde\phi$.
In order to deduce an identity on the integrand of (\ref{eq-R-temp}), we first need to integrate it by parts and symmetrize it with respect to $t$ and $t'$. This yields two equations that are in fact redundant, and hence we deduce the following sufficient condition for $R_1$ and $R_2$:
\begin{align}
    R_1 (t) - R_1 (-t) + \dot R_2 (t) -\dot R_2 (-t) = 0
\end{align}

\section{Flow equations }
\label{sec_ONflow}

One can show for the model A that the dynamical parts of the dimensionless renormalization functions read 
\footnote{Notice that our equations differ from those of reference~\cite{Canet11} that involve a missprint.}:
\begin{align}
\label{eq:flowup}
    &\partial_t \hat{U}' |_{\text{dyn}} =  - v_d \int_y y^{d/2} \frac{f s}{h^2} \\
    \begin{split}
    \label{eq:flowz}
    &\partial_t \hat{Z} |_{\text{dyn}} = 2 v_d \int_y y^{d/2} \frac{s}{h^2}\left[ 
    \frac{2 \hat\rho f^2}{h^2} \left(\frac{4}{d}\frac{yh'^2}{h}-h'-\frac{2}{d}yh'' \right)  \right. \\
    & \left. 
    + 4 \hat\rho \hat{Z}'\frac{f}{h} \left(1-\frac{2}{d}\frac{yh'}{h} \right)
    + \frac{2 \hat\rho}{d} (\hat{Z}')^2 \frac{y}{h} - \frac{\hat{Z}'}{2}
    -\hat\rho \hat{Z}''
    \right] 
    \end{split}
    \\
    &\partial_t \hat{X} |_{\text{dyn}} =  v_d \!\! \int_y y^{d/2} \frac{s}{h^2} \!\! \left(8\hat\rho \hat{X}' \frac{f}{h} \!
    -3\frac{f^2}{h^2} \hat\rho \hat{X} \! -\hat{X}' \! - 2\hat\rho \hat{X}''  \right)
    \label{eq:flowx}
\end{align}
where $\partial_t \equiv k \partial_k$, $v_d^{-1}= 2^{d+1}\pi^{d/2}\Gamma(d/2)$, $\hat\rho=\hat\psi^2/2$,  $h(y,\hat\rho)=y(\hat{Z}(\hat\rho)+r(y))+\hat{U}'(\hat\rho)+2\hat\rho \hat{U}''(\hat\rho)$, $f(y,\hat\rho)=y\hat{Z}'(\hat\rho)+3\hat{U}''(\hat\rho)+2\hat\rho \hat{U}'''(\hat\rho)$ and $s(y)=-\eta r(y)-2yr'(y)$. One notices immediately that $\hat{X}$ does not contribute to the flows of $\hat{U}'$ and $\hat{Z}$, that are the standard flows of the static Ising model.

For the dynamical $O(N)$ model, for simplicity we only consider the Local Potential Approximation prime (LPA') of the Derivative Expansion, which means we only retain $U'_k$ as a function of $\rho$, and $Z_k$ and $X_k$ are mere numbers. While the flow of the dimensional part of the dimensionless renormalization functions is still given by Eqs.~(\ref{eq:dimPartu})-(\ref{eq:dimPartx}), the flow of the dynamical part is given this time by the following equations:
\begin{align}
       \partial_t \hat{U}' |_{\text{dyn}} &=  - v_d \int_y y^{d/2} s  \left(\frac{3\hat{U}''+2\hat \rho \hat{U}^{(3)}}{h_L^2} + \frac{(N-1)\hat{U}''}{h_T^2} \right) \label{eq:dynON1}\\
    \begin{split}
        \partial_t \hat{Z} |_{\text{dyn}} &= -8 v_d \int_y y^{d/2} s \hat \rho \hat{U}'' \left( \frac{h_y}{h_L^2h_T^2} +\right. \\
        & \left. \left. +\frac{2y h_{yy}}{d h_L^2h_T^2}-\frac{2y h_y^2}{d h_L^2h_T^2}\left(\frac{1}{h_T}+\frac{1}{h_L} \right) \right) \right|_{\hat \rho=\hat \rho_{0,k}}\label{eq:dynON2}
    \end{split} \\
        \partial_t \hat{X} |_{\text{dyn}} &= \left. -4 v_d \! \! \int_y \! y^{d/2} s \hat \rho \hat{U}''^2 \! \left( \! \frac{h_L^2+4h_L h_T + h_T^2}{h_L^2 h_T^2 (h_L+h_T)^2} \! \right)
        \right|_{\hat \rho=\hat \rho_{0,k}}\label{eq:dynON3}
\end{align}
where $s(y)=-\eta r(y)-2yr'(y)$, $h_L=y (r(y)+1)+\hat{U}'+2\hat \rho \hat{U}''$, $h_T=y (r(y)+1)+\hat{U}'$,
$h_y = 1+r(y)+r'(y)$ and $h_{yy} = 2r'(y)+y r''(y)$. Notice that since we are working at the LPA', the flow equations for $\hat{Z}$ and $\hat{X}$ are evaluated at the (running) minimum of the potential $\rho_{0,k}$. Once again, $\hat{X}$ does not contribute to the flows of $\hat{U}'$ and $\hat{Z}$, that are the standard flows of the equilibrium $O(N)$ model at the LPA'.

\section{Causality and Kramers-Kronig theorem}
\label{sect:KK}

The linear response function $\chi (t,t')$ is defined to be the variation of the mean value of the field $\phi$ at time $t$ caused by the variation of the external source $J$ coupled to $\phi$ at time $t'$. Mathematically, it reads:
\begin{align}
    \chi (t,t') &= \frac{\bra \delta \phi (t) \ket}{\delta J(t')}|_{J\to 0}\, .
\end{align}
Because of time translation invariance, it is a function of $t-t'$ only and we may write $\chi(t,t')=\chi (t-t')$. In the MSRDJ formalism (also called response-function formalism), the response-function reads: 
\begin{align}
    \chi (t,t') &= \bra \tilde\phi (t') \phi (t) \ket \, ,
\end{align}
and its Fourier Transform $\chi (\omega)$ is simply given by the upper-right element of the propagator matrix $G_k$:
\begin{align}
    \chi (\omega) &= \frac{1}{P(q^2,-\omega)} = \frac{1}{h(q^2,-\omega)-i \omega X_k} \, ,
\end{align}
with $h(q^2,\omega) = Z_k(\rho)q^2+R_{1,k}(q^2,\omega)+U'_k(\rho)+2\rho U''_k(\rho)$. Causality imposes $\chi (t<0)=0$, which means that $\chi(\omega)$ must be an analytic function of $\omega$ in the upper-part of the complex plane. In other words, the poles of $\chi (\omega)$ must have a negative imaginary part.

\pagebreak

\onecolumngrid

Let us add that the Kramers-Kronig theorem provides an alternative translation of the causality of the response function. Indeed, the fact that $\chi (t<0)=0$ yields the following equalities for the Fourier Transform $\chi (\omega)$, called the Kramers-Kronig relations~\cite{Tauber_book}:  
\begin{align}
    \text{Re}(\chi(\omega)) &= \frac{1}{\pi} \mathcal{P} \int \dd \omega' \, \frac{\text{Im}(\chi (\omega'))}{\omega'-\omega} \\
    \text{Im}(\chi(\omega)) &= - \frac{1}{\pi} \mathcal{P} \int \dd \omega' \, \frac{\text{Re}(\chi (\omega'))}{\omega'-\omega}
\end{align}
where $\mathcal{P}$ denotes the Cauchy principal value of the integral. 

\twocolumngrid

\bibliographystyle{apsrev4-1} % Tell bibtex which bibliography style to use
\bibliography{PRL_Langevin/bibliotheque} % Tell bibtex which .bib file to use 
 
\end{document}